\documentstyle{aipproc}
\input psfig.tex
\begin{document}
\title{  The GZK Bound and Strong Neutrino-Nucleon 
 Interactions  above $10^{19}$ eV: a Progress Report}

\author{ John P. Ralston $^a$, Pankaj Jain$^b$,\\
 Douglas W. McKay$^a$, and S. Panda$^b$ }
\address{
$^a$	Department of Physics \& Astronomy\\
	University of Kansas,
	Lawerence, KS 66045, USA\\
$^b$              Physics Department,
	I.I.T. Kanpur, India 208016}
\maketitle

\begin{abstract}
 Cosmic ray events above $10^{19}$ eV have posed a
fundamental problem for more than thirty years.  Recent measurements
indicate that these events do not show the features predicted by the
GZK bound. The events may, in addition, display angular correlations
with points sources.  If these observations are confirmed for point
sources further than 50-100 Mpc, then strong interactions for the
ultra-high energy neutrino are indicated.  Recent work on extra
space-time dimensions provides a context for massive spin-2
exchanges which are capable of generating cross sections in the
$1-100 $ mb range indicated by data.  Some recent controversies on the
applicability of extra-dimension physics are discussed.
\end{abstract}

\section{\it  The GZK Puzzle}
For more than 35 years it has been known
extra-galactic proton cosmic rays should not exceed the energy of
about $5\times 10^{19}$ eV. This is the GZK bound \cite{GZK}.  Year after
year, observations have continued to defy this result \cite{linsley}.  Recent
observations of showers with reliably determined energies
above $10^{20} $ eV \cite {fliesI} deepen the puzzle.

The GZK bound is as basic as the Big Bang and the Standard Model.  The
bound is obtained by attenuation of ultra-high energy (UHE) protons on
the cosmic microwave background.  The threshold of $5\times 10^{19}$ eV is
kinematic, representing a center of mass energy sufficient for
production of electron-positron pairs and nuclear resonances.  The
cross sections for these reactions and the density of
microwave background photons are known.  So the attenuation of protons
is known, and exponential attenuation will eliminate protons from sources more
about 50 Mpc from the Earth.  There are not enough sources within 50 Mpc
to explain the data, making the origin of the $GZK$-violating
events very mystifying.  Speculative models such as unstable
superheavy relic particles and topological defects
can create particles of high enough energy, but also
require sources within the same limited sphere.  A hint of
correlations between observed event directions and cosmologically-distant
sources \cite{Farrar98}, if confirmed, also indicates something very
mysterious.

The only elementary particle that could
cross the requisite intergalactic distances of about 100 Mpc or more
is the neutrino.  The only electrically neutral, stable particle that
could expain correlations with cosmologically-distant sources is also
the neutrino.  Flux estimates of $UHE$ neutrinos produced by extra-galactic
sources and GZK-attenuated nucleons and nuclei vary widely, but suffice to
account for the shower rates observed. The neutrino would be a 
natural candidate
for the events, if its interaction cross section were large enough
\cite{elbert}.

The neutrino-nucleon total cross section $\sigma_{tot}$ is the
crucial issue. In the Standard Model $\sigma_{tot}$ is based
on small-x QCD evolution and $W^{\pm}, Z$ exchange physics.
Cross sections of order $10^{-4}\ -\ 10^{-5}$ mb are predicted for 
the region of $10^{20}$ eV
primary energy. This is far too small to explain the data: so perhaps 
new physical processes may be at work.

\subsection{\it The $\nu N$ Cross Section with Massive Spin-2 Exchange}

Total cross sections at high energies are dominated by characteristics
of the t-channel exchanges.  The growth of $\sigma_{tot}$ with energy,
in turn, is directly correlated with the {\it spin} of exchanged
particles.

For this reason the GZK-puzzle cannot be explained by considering new
spin-1 exchanges.  New  W- or Z-like massive {\it vector}
bosons would produce $\sigma_{tot}$ growing at the same rate as the
standard one, and normalized below it.  Indeed any new neutrino physics must
not disturb the agreement of laboratory data with 4-Fermi physics in
the low energy region.  At the same time the $GZK$-violating data indicates
a cross section growing with energy that must greatly exceed the Standard
Model growth by  $10^{20}$ eV. How is this possible?

The paradox is beautifully resolved with {\it massive spin-2}
exchange.  The fascinating possibility of massive spin-2 exchange has
recently become popular in the context of hidden ``extra"
dimensions\cite{ADD1}.  Kaluza-Klein ($KK$) excitations of the
graviton are believed to act like a tower of massive spin-2 particle
exchanges.  We calculate the effect using the Feynman rules
developed by several groups\cite{HaLyZh}. We find neutrino-nucleon
cross section values at $E_{\nu}=10^{20}$ eV in the $1-100$ mb
range from mass scales in the $1-10$ TeV range.  Interestingly,
this range of masses is just that indicated by the extra-dimensions
phenomenology, and this range of cross sections is just that
indicated by the $GZK$-violating data.

What about low energies? With spin-2 exchange, $d\sigma/dt$ grows like $s^2$
and $\sigma$ like $s^3 $ in the low-energy, perturbative region (with a 
cutoff on t, then one power of s in the total cross section is replaced by
the cutoff, of course).
The new contributions to $\sigma_{tot}$ are naturally orders of
magnitude below the Standard Model component in the entire regime
where neutrino cross sections have been measured.  Nor is any miracle
needed for the neutrino interactions to emerge above ordinary cosmic
ray events above $10^ { 19 }$ eV.  That is because the $GZK$ attenuation process
must set in, uncovering a spectrum of neutrinos just at the
point where protons get attenuated  away.  Our model of neutrinos and 
massive spin-2 exchange
explain the $GZK$-violating events consistent with all known
experimental limits, as shown in Fig. 1.

\begin{figure}
\psfig{file=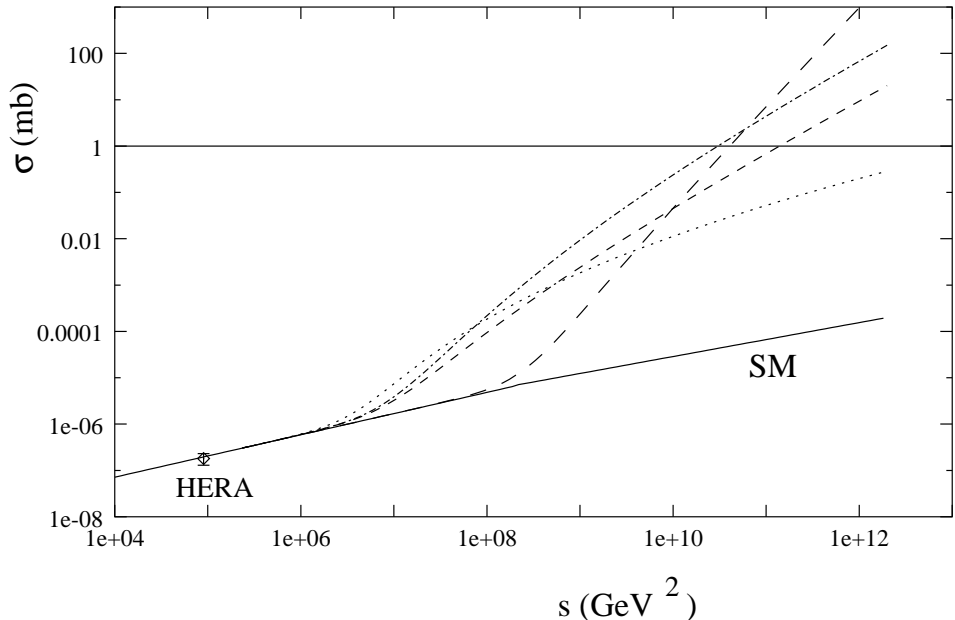}
\caption{The $\nu N$ cross section in the Standard Model ($SM$)
compared to a theory with large extra dimensions and three different
models for the unitarity extrapolation between perturbative to non-
perturbative regimes.  The dotted line shows the $log(s)$ growth case with
$M_{S}$ = $1$ TeV and $\xi$ = 10.  The short dashed and dash-dotted lines
show $s^1$ growth with $M_{S} = 1$ TeV and $\beta = 1$ and 0.1
respectively.  The long
dashed line shows $s^2$ growth with $M_{S} = 3$ TeV and $\beta = 1$.  The
contribution from
massive graviton exchange is negligible at low energies but rises
above the $SM$ contribution when $\sqrt s > M_{S}$, reaching typical
 hadronic cross sections
at incident neutrino energies in the range $5\times10^{19}$ to
 $5\times10^{20}$ GeV.
The HERA data point is shown for comparison.
The approximate minimum value required for $\nu N$ cross-section,
 $\sigma=1$ mb, is indicated by the horizontal straight line.}

\end{figure}

\subsection{\it  Updates and Controversies}

This is an exciting area with rapid progress on many applications.

Striking signatures appear in predictions of signals in planned
$Km^3$-scale cosmic neutrino detectors.  The 1 TeV to 10 PeV region
will be explored,  much like the existing AMANDA\cite{amanda}
and RICE\cite{rice} detectors.  In this regime the predicted 
\cite{us} ratio of upward to downward neutrino-induced showers 
differs substantially from
the Standard Model, creating a nice diagnostic for new physics.

Horizontal air showers develop over long distances and probe smaller
cross sections.  
Using essentially our
enhanced cross section with $s^1$ behaviour, but with different
parameter choices, the authors of Ref. \cite{tos} link horizontal
shower studies to extra dimension physics. They
show how fundamental scales up to 10 TeV can be probed.  Detailed
studies of  shower characteristics generated by our model for
various cross sections are also underway\cite{pankaj}. We find that 
the neutrino-induced and proton-induced showers cannot be
distinguished on a shower-by-shower basis.  There are exciting prospects
for finding signals of new physics in shower angular distributions.

A recent re-calculation within the same Kaluza-Klein $KK$
framework \cite{kp} includes the ``brane recoil'' effects advocated
by\cite{bando}.  
After making several technical points, the authors of Ref. \cite{kp} present a
cross section that is basically equivalent to our $s^2$ growth model.  
They conclude that the resulting cross section is too small to explain
the UHE cosmic ray showers. However they do not consider all the parameter
choices allowed by current experimental limits and hence we find their
conclusions unsupported.

{\bf Acknowledgments:} We thank Tom Weiler, Prashanta Das, Faheem Hussain,
and Sreerup Raychaudhuri for useful discussions.  This work
was supported in part by U.S. DOE Grant number DE-FG03-98ER41079 and
the {\it Kansas Institute for Theoretical and Computational Science}.

\end{document}